\documentclass[a4paper,10pt]{article}
\usepackage[english]{babel}
\usepackage{float}
\usepackage{amsmath}
\usepackage{graphicx,float}
\usepackage{authblk}
\usepackage{hyperref}

\usepackage{xcolor}
\usepackage{tipa}
\usepackage{ulem}

\begin{document}

\title{Photon-pion transition form factor\\ of two photon process at BESIII \\from light-cone quantum field theory\thanks{published in Nucl.Phys.A 1027 (2022) 122497,
\url{https://doi.org/10.1016/j.nuclphysa.2022.122497}}}


\author[1]{Yining Xian}
\author[1,2,3]{Bo-Qiang Ma\footnote{Email address: mabq@pku.edu.cn}}
\affil[1]{School of Physics, Peking University, Beijing
100871, China}
\affil[2]{Center for High Energy Physics, Peking University, Beijing 100871, China}
\affil[3]{Collaborative Innovation Center of Quantum Matter, Beijing, China}

\date{}

\newcommand\keywords[1]{\textbf{Keywords}: #1}

\maketitle

\begin{abstract}
This work studies the experiments of two-photon processes at BESIII from a theoretical viewpoint in light-cone quantum field theory. We adopt the momentum space wave function of the pion in the light-cone formalism by two forms of the Brodsky-Huang-Lepage~(BHL) prescription and the light-cone holographic QCD. We find that the wave function got by the light-cone holographic QCD can be related to the BHL wave function by a Jacobian factor. Then we calculate the transition form factor~(TFF) of the two-photon process by these two wave functions under the light-cone quark model. We compare the calculated results with the preliminary experimental data at BESIII and also other theoretical calculations.

\end{abstract}



\section{Introduction}

Two-photon processes, including Compton scattering, two-photon annihilation to a particle and anti-particle pair, the process of Higgs to two-photons, and photons in high-energy cosmic rays etc., have been studied for a long time.  These experiments can examine chiral symmetry, quark mass ratios, mixing and decay of light pseudoscalar mesons and some QCD-based physical models. We mainly study the transition form factor~(TFF) of the two-photon to pseudoscalar meson process. At the BESIII experiment in Beijing, the transition form factors of
pseudoscalar mesons are studied in their Dalitz decays~\cite{BESIII:2015zpz}, their radiative
production in $e^{+}e^{-}$ annihilation, and their production in two-photon
scattering. BESIII carries out these experiments of two photons to pseudoscalar mesons($\pi$, $\eta$, $\eta^{\prime}$) in order to study the anomalous magnetic moment of $\mu$, and obtain a preliminary data of photon-pion transition form factor at $Q^2=0\to 3~\mathrm{GeV}^2$~\cite{Redmer:2017fhg,Redmer:2019iye}. There is a large amount of $e^{+}e^{-}$ collision data for BESIII near this energy region~\cite{Ablikim:2013ntc,BESIII:2015qfd,BESIII:2016kpv,BESIII:2017tvm}. The transition form factors at these energies are the mainly input for the hadronic light-by-light contribution to the muon $(g-2)$. Although there have been many meson TFF experiments before~\cite{CELLO:1990klc,CLEO:1997fho,BaBar:2009rrj,Belle:2012wwz}, the previously experimental data are very little and are not 
precise at this energy scale. The latest preliminary results of BESIII have a better accuracy compared with the previous experiments. Perturbative QCD is not applicable at this energy scale, here we use the light-cone quantum field theory to calculate the photon-pion transition form factor and make a comparison with these data.

Light-cone quantum field theory uses a light-cone form which is first proposed by Dirac to parameterize spacetime~\cite{Dirac:1949cp}. Its advantage is that the structure of vacuum is relatively simple. Hadrons have a Fock-state expansion in this framework. By using this expansion and the light-cone sum rule~\cite{Brodsky:1997de}, various properties of hadrons can be calculated.

During the last decade, a connection is established between the conformal quantum mechanics and the light-cone dynamics~\cite{Brodsky:2003px}, which provides a natural framework to reconcile the quark-parton model with QCD. Therefore, the AdS/CFT correspondence can be used to obtain an approximation to QCD. The light-cone holographic QCD provides a new method to solve the strong interaction problem from basic assumptions. It has led to many remarkable results~\cite{Brodsky:2014yha}. We can get momentum space wave function in the Fock-state expansion of the hadron by the light-cone holography mapping.

In the calculation of TFF under the light-cone quark model, it is necessary to obtain the momentum space wave function. A common method is to make an approximation in the instant form and then convert to the light-cone form. We found that if we consider the Jacobian factor caused by a conversion from a non-relativistic form to a relativistic form, the momentum wave function obtained in this way is the same as the wave function obtained by the soft-wall model of the light-cone holography method.

We also review the theoretical constraints on the valence wave function, and use more reasonable constraints to limit the parameters in the momentum wave function and calculate the TFF. The calculated results are in good agreement with the preliminary results of BESIII below $Q=1~\mathrm{GeV}$.

In Sec.~2 we briefly introduce the relevant experiments at BESIII. In Sec.~3, we obtain specific expression of transition form factor in light-cone formalism. In Sec.~4, we obtain the momentum space wave function in two ways. In Sec.~5, we calculate the transition form factors of $ \gamma \gamma^{*}\rightarrow\pi$ and compare it with the experiment data at BESIII. We compare our results with the lattice QCD calculation and other results in Sec.~6. In Sec.~7, we present a brief summary.

\section{Experiment at BESIII}

   Transition form factor~(TFF) describes the coupling of photons with hadronic matter. The large data acquired at BESIII~\cite{Ablikim:2013ntc,BESIII:2015qfd,BESIII:2016kpv,BESIII:2017tvm} allow people to study the transition form factors~(TFF) of pseudoscalar mesons. The two-photon physics program of the BESIII Collaboration is mainly motivated by the need of new measurements of transition form factors as input for the Standard Model~(SM) calculations of the anomalous magnetic moment of the muon~\cite{Redmer:2019iye}.
   
The SM prediction of the anomalous magnetic moment of the muon is dominated by the QED contribution, which has been calculated up to 5-loop in perturbative theory, with a precision of 0.0007~ppm~\cite{Aoyama:2012wk}. The weak contribution which is very small has been calculated to 2-loop~\cite{Gnendiger:2013pva}. The current limitation of the precision of the SM calculation is the hadronic contribution. This contribution has two parts, one part is hadronic vacuum polarization~(HVP) contribution, the other part is hadronic light-by-light (HLbL) contribution. The valid model for calculating hadronic light-by-light (HLbL) contribution needs the meson transition form factor (TFF) as an input.

The BESIII detector~\cite{BESIII:2009fln} is a magnetic spectrometer located at the Beijing Electron Position Collider (BEPCII). The BESIII experiment has collected a set of data samples at center-of-mass (CM) energies from 2.0~GeV to 4.6~GeV ~\cite{Ablikim:2013ntc,BESIII:2015qfd,BESIII:2016kpv,BESIII:2017tvm} which covers the whole energy region used for measurements of $R$ ratio, $\tau$ physics quantities, and relevant form factors.
TFF can be measured with different techniques in different kinematic regions. Dalitz decay of light pseudoscalar mesons provides access to the time-like TFF of the decaying mesons~\cite{BESIII:2022cul}. The photon-pion transition form factors discussed in this paper are mainly measured with $e^+e^-$ colliders.
Each of the leptons emits a photon, and mesons are the production in two-photon collisions, as shown in Fig.~1. 
The cross section of meson production in two-photon collisions is directly proportional to the square of the TFF in the space-like region. The momentum dependence of the TFF can be studied from the momentum transfer of the scattered leptons~\cite{Guo:2019gjf}.

 \begin{figure}[H]
    \centering
    \includegraphics[width=0.8\textwidth]{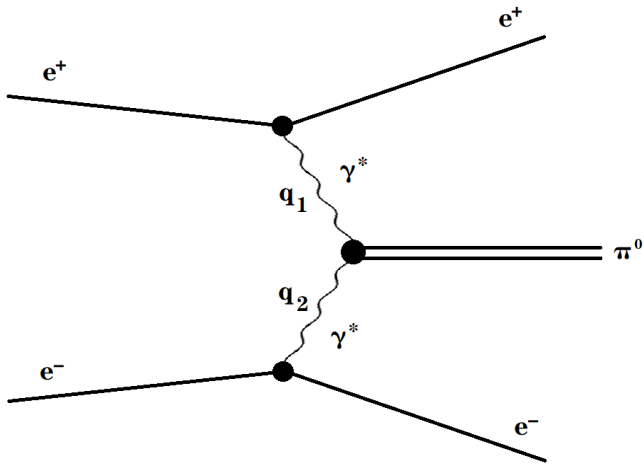}
    \caption{The diagram for the two-photon fusion process.}
    \label{fig:my_label}
\end{figure}

 There are three types of meson TFF that can be measured through the two-photon process depending on the number of leptons received by the detector~\cite{Guo:2019gjf}. In the untagged case, only hadronic productions are detected. By requiring both leptons are parallel with the beam directions, the virtuality of both photons is very small~($q_{1,2}^{2} \simeq 0$), and they can be considered as quasi-real. In the single tagged case, one of the final leptons is detected in the detector, while the other is required to be scattered along the beam direction. In this case, the photon emitted from the tagged lepton is far off-shell with $q_{1}^{2}=-Q^{2}$, while the untagged one is quasi-real, with $q_{2}^{2} \simeq 0$. The TFF as a function of $Q^{2}$, $F_{\gamma^{*}\gamma^{*}\rightarrow\pi}\left(q_{1}^{2}, q_{2}^{2}\right) \rightarrow F_{ \gamma \gamma^{*}\rightarrow \pi }\left(Q^{2}\right)$ can be measured. In the double tagged case, all final states are detected, and the TFF $F_{   \gamma^{*} \gamma^{*}\rightarrow\pi}\left(q_{1}^{2},q_{2}^{2}\right)$ is accessible. The double tagged method is limited by statistics as the cross-section of the two-photon process strongly peaks at small angle, so most of the BESIII current measurements are done with untagged or single tagged methods.

BaBar and Belle measured the transition form factor in the space-like region in the past~\cite{BaBar:2009rrj,Belle:2012wwz}. CELLO and CLEO also performed related experiments in the 1990s~\cite{CELLO:1990klc,CLEO:1997fho}. These measurements are given at energy scales above 4 GeV with higher precision. However, these experiments have less experimental data below 3 GeV, only CELLO's experimental data give some data below 1.5 GeV, and the precision of these data is not precise enough. The part where TFF has the greatest impact on HLbL comes from below 1 GeV. 
The BESIII experiment runs at much lower C.M. energies than the B-factories, and thus has the advantage in the measurements of the most relevant energy region.

\section{Formalism of photon-pion transition form factor}
In the single tagged case, the quasi-real photon can be considered as an on-shell photon. The transition form factor, in which an on-shell photon is scattered by one off-shell photon and decays into a meson, as schematically shown in Fig. 2, is defined by the 
$ \gamma \gamma^{*}\pi^{0}$ vertex
$$
\Gamma_{\mu}=-i e^{2} F_{\gamma\gamma^{*}  \rightarrow \pi^{0}}\left(Q^{2}\right) \varepsilon_{\mu \nu \rho \sigma} p_{\pi}^{\nu} \epsilon^{\rho} q^{\sigma},
$$
in which $q$ is the momentum of the off-shell photon, $-Q^{2}=q^{2}=q^{+} q^{-}-\mathbf{q}_{\perp}^{2}=-\mathbf{q}_{\perp}^{2}$ and $\epsilon$ is the polarization vector of the on-shell photon, and 
$$\Gamma^{+}=\left\langle\Psi_{\gamma}^{\uparrow}\left(P^{+}, \mathbf{P}_{\perp}\right)\left|J^{+}\right| \Psi_{\pi}\left(P^{\prime+}, \mathbf{P}_{\perp}^{\prime}\right)\right\rangle \delta^{3}\left(\mathbf{P}+\mathbf{q}-\mathbf{P}^{\prime}\right).$$
It can be calculated in light-cone formalism with the Fock expansion and light-cone sum rule. 
 \begin{figure}[H]
    \centering
    \includegraphics[width=0.8\textwidth]{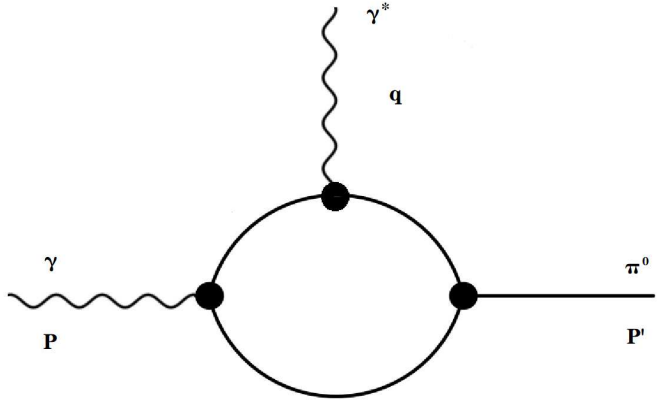}
    \caption{The diagram for the contribution to the TFF $ F_{\gamma \gamma^{*} \rightarrow \pi^{0}}$. }
    \label{fig:my_label2}
\end{figure}
The light-cone formalism has a simple vacuum which we can generate a complete set of Fock-states~\cite{Brodsky:1997de}. A hadron can be expanded in Fock-state basis as
$$
|H\rangle=\sum_{n}  \Psi_{n / H}\left(x_{i}, \mathbf{k}_{\perp i}, \lambda_{i}\right)\left|n: k_{i}^{+}, \mathbf{k}_{\perp i}, \lambda_{i}\right\rangle.
$$
for example, for the pion
$$
|\pi\rangle=\sum|q \bar{q}\rangle \Psi_{q \bar{q}}+\sum|q \bar{q} g\rangle \Psi_{q \bar{q} g}+\cdots,
$$
and the temporal evolution of the state is generated by the light-cone Hamiltonian $\mathrm{H}_{\mathrm{LC}}^{\mathrm{QCD}} $.
Similar with pion, one can assume that the photon may also have this kind of Fock-state expansion~\cite{Xiao:2003wf} in the process shown in Fig.~2
$$
|\gamma\rangle=\sum|q \bar{q}\rangle \Psi_{q \bar{q}}.
$$

The energy region we consider here is below 3~GeV, and the TFF is only related to the valence state, so we only take into
account the minimal Fock-states of the pion and the photon.
$\Psi$ is the amplitude of the hadron state component, also called the hadron wave function, which can be divided into a spin part and a momentum part 
$$
\Psi\left(x_{i}, \mathbf{k}_{\perp i}, \lambda_{i}\right)=\psi\left(x_{i}, \mathbf{k}_{\perp i}\right) \chi\left(x_{i}, \mathbf{k}_{\perp i}, \lambda_{i}\right).
$$

Spin wave function of the pion is got in paper~\cite{Xiao:2003wf} by a fully relativistic field theory treatment of the interaction vertex along with the idea in~\cite{Brodsky:1980zm,Brodsky:2000ii}. In those papers it is assumed that one can consider the pion vertex connecting to two $\text { spin- } \frac{1}{2}$ fermions (e.g. two quarks) by only taking into account the minimal Fock-state. We can obtain the above four components of the spin wave function by calculating the matrix elements of
$$
\frac{\bar{u}\left(k_{1}^{+}, k_{1}^{-}, \mathbf{k}_{\perp}\right)}{\sqrt{k_{1}^{+}}} \gamma_{5} \frac{v\left(k_{2}^{+}, k_{2}^{-},-\mathbf{k}_{\perp}\right)}{\sqrt{k_{2}^{+}}},
$$
and then

$$
\begin{aligned}
\left\langle\Psi_{\pi}\left(P^{+}, \mathbf{P}_{\perp}\right)\right|=& \int \frac{\mathrm{d}^{2} \mathbf{k}_{\perp} \mathrm{d} x}{16 \pi^{3}} \\
& \times\left[\Psi_{\pi L}\left(x, \mathbf{k}_{\perp}, \uparrow, \downarrow\right)\left\langle x P^{+}, \mathbf{k}_{\perp}, \uparrow, \downarrow\right|+\Psi_{\pi L}\left(x, \mathbf{k}_{\perp}, \downarrow, \uparrow\right)\left\langle x P^{+}, \mathbf{k}_{\perp}, \downarrow, \uparrow\right|\right.\\
&\left.+\Psi_{\pi L}\left(x, \mathbf{k}_{\perp}, \uparrow, \uparrow\right)\left\langle x P^{+}, \mathbf{k}_{\perp}, \uparrow, \uparrow\right|+\Psi_{\pi L}\left(x, \mathbf{k}_{\perp}, \downarrow, \downarrow\right)\left\langle x P^{+}, \mathbf{k}_{\perp}, \downarrow, \downarrow\right|\right],
\end{aligned} 
$$
in which,
$$
\Psi_{\pi L}\left(x, \mathbf{k}_{\perp}, \uparrow, \downarrow\right)=-\frac{m}{\sqrt{2\left(m^{2}+\mathbf{k}_{\perp}^{2}\right)}} \psi_{\pi}^{*} ;\\
$$
$$
\Psi_{\pi L}\left(x, \mathbf{k}_{\perp}, \downarrow, \uparrow\right)=+\frac{m}{\sqrt{2\left(m^{2}+\mathbf{k}_{\perp}^{2}\right)}} \psi_{\pi}^{*}; \\
$$
$$
\Psi_{\pi L}\left(x, \mathbf{k}_{\perp}, \uparrow, \uparrow\right)=+\frac{k_{1}+i k_{2}}{\sqrt{2\left(m^{2}+\mathbf{k}_{\perp}^{2}\right)}} \psi_{\pi}^{*}; \\
$$
$$
\Psi_{\pi L}\left(x, \mathbf{k}_{\perp}, \downarrow, \downarrow\right)=+\frac{k_{1}-i k_{2}}{\sqrt{2\left(m^{2}+\mathbf{k}_{\perp}^{2}\right)}} \psi_{\pi}^{*},
$$
where $\psi_{\pi}^{*}$ is the momentum space wave function.

Similar to the pion vertex, paper~\cite{Xiao:2003wf} obtains the spin
wave function of the spin-1 photon from the minimal Fock-state basis by calculating the matrix elements of
$$\frac{\bar{v}\left(k_{2}^{+}, k_{2}^{-}, \mathbf{k}_{ \perp}\right)}{\sqrt{k_{2}^{+}}} \gamma \cdot \epsilon^{*} \frac{u\left(k_{1}^{+}, k_{1}^{-}, -\mathbf{k}_{\perp }\right)}{\sqrt{k_{1}^{+}}},
$$
with results

$$
\Psi_{L}^{\uparrow}\left(x, \mathbf{k}_{\perp}, \uparrow, \downarrow\right)=-\frac{\sqrt{2}\left(k_{1}-i k_{2}\right)}{1-x} \psi_{\gamma}, {\left[l^{z}=+1\right]};
$$

$$  \\
\Psi_{L}^{\uparrow}\left(x, \mathbf{k}_{\perp}, \downarrow, \uparrow\right)=+\frac{\sqrt{2}\left(k_{1}-i k_{2}\right)}{x} \psi_{\gamma}, {\left[l^{z}=+1\right]};
$$

$$
 \\
\Psi_{L}^{\uparrow}\left(x, \mathbf{k}_{\perp}, \uparrow, \uparrow\right)=-\frac{\sqrt{2} m}{x(1-x)} \psi_{\gamma}, {\left[l^{z}=0\right]};
$$

$$
 \\
\Psi_{L}^{\uparrow}\left(x, \mathbf{k}_{\perp}, \downarrow, \downarrow\right)=0, 
$$
in which,
$$
\psi_{\gamma}=\frac{e_{q}}{\lambda^{2}-\frac{m^{2}+\mathbf{k}_{\perp}^{2}}{x}-\frac{m^{2}+\mathbf{k}_{\perp}^{2}}{1-x}},
$$
where $\lambda$ is the photon mass and equals to 0. Therefore, the quark-antiquark Fock-state for the photon is given by
$$
\begin{aligned}
\left\langle\Psi_{\gamma}^{\uparrow}\left(P^{\prime+}, \mathbf{P}_{\perp}^{\prime}\right)\right|=& \int \frac{\mathrm{d}^{2} \mathbf{k}_{\perp} \mathrm{d} x}{16 \pi^{3}} \\
& \times\left[\Psi_{L}^{\uparrow}\left(x, \mathbf{k}_{\perp}, \uparrow, \downarrow\right)\left\langle x P^{\prime+}, \mathbf{k}_{\perp}, \uparrow, \downarrow\right|+\Psi_{L}^{\uparrow}\left(x, \mathbf{k}_{\perp}, \downarrow, \uparrow\right)\left\langle x P^{\prime+}, \mathbf{k}_{\perp}, \downarrow, \uparrow\right|\right.\\
&\left.+\Psi_{L}^{\uparrow}\left(x, \mathbf{k}_{\perp}, \uparrow, \uparrow\right)\left\langle x P^{\prime+}, \mathbf{k}_{\perp}, \uparrow, \uparrow\right|+\Psi_{L}^{\uparrow}\left(x, \mathbf{k}_{\perp}, \downarrow, \downarrow\right)\left\langle x P^{\prime+}, \mathbf{k}_{\perp}, \downarrow, \downarrow\right|\right].
\end{aligned}
$$
With color and flavor considered, paper~\cite{Xiao:2003wf} uses these expansion and light-cone sum rules to give the specific form of the TFF $F_{\gamma \gamma^{*} \rightarrow \pi^{0}}\left(Q^{2}\right)$ as
$$
F_{\gamma \gamma^{*} \rightarrow \pi}\left(Q^{2}\right)=4 \sqrt{3}\left(e_{u}^{2}-e_{d}^{2}\right) \int_{0}^{1} \mathrm{~d} x \int \frac{\mathrm{d}^{2} \mathbf{k}_{\perp}}{16 \pi^{3}} \psi_{\pi}\left(x, \mathbf{k}_{\perp}^{\prime}\right) \frac{m}{x \sqrt{m^{2}+\mathbf{k}_{\perp}^{\prime 2}}} \frac{x(1-x)}{m^{2}+\mathbf{k}_{\perp}^{2}},
$$
in which $\mathbf{k}_{\perp}^{\prime}=\mathbf{k}_{\perp}+(1-x) \mathbf{q}_{\perp}$, $\psi_{\pi}\left(x, \mathbf{k}_{\perp}^{\prime}\right)$ is the momentum space wave function. 

\section{The momentum space wave function}
From the discussion above, it can be known that the spin wave function is relatively certain in the calculation of the hadronic properties in the light-cone model. Momentum space wave function contains kinetic information related to non-perturbative QCD in hadrons. The largest uncertainty comes from the choice of the momentum space wave function. The exact hadron wave function needs to be obtained by solving the BS equation~\cite{Schwinger:1951ex}. It is very difficult to solve the BS equation exactly in numerical calculations, so some approximate forms are proposed to simplify the calculation. This work uses two methods to obtain the momentum space wave function, one is the Brodsky-Huang-Lepage~(BHL) prescription~\cite{Brodsky:1982nx,Huang:1994dy}, and 
the other is the light-cone holographic approach. The BHL prescription uses a simply assumption which has been proved by many works~\cite{Liu:2014npa,Xiao:2003wf,Qian:2008px} agreeing well with the experiment data. The light-cone holographic QCD can derive the momentum space wave function from a first approximation. 
Liu and Ma have used this model to calculate baryon form factors and got  results compatible with data~\cite{Liu:2015jna}. 

\subsection{The BHL prescription}
The Brodsky-Huang-Lepage~(BHL) prescription~\cite{Brodsky:1982nx,Huang:1994dy}  uses the harmonic oscillator approximation, and then converts the instant harmonic oscillator wave function to the light-cone frame. The ground-state momentum space wave function of the harmonic oscillator potential in the instant form is given by
$$
\psi_{0}(\mathbf{q}) \propto e^{-\frac{1}{2 \beta^{2}} \mathbf{q}^{2}}.
$$
In the case of two-particle system with $m_1=m_2=m$, the momentum relationship between the light-cone
frame and the instant frame is~\cite{Brodsky:1982nx,Huang:1994dy} 
\begin{equation*}
\mathbf{q}^{2} =\frac{\mathbf{k}_{\perp }^{2}+m^{2} }{4x(1-x)}-m^2 .
\end{equation*}
Then we can get
$$
\psi_{\pi}(x, \mathbf{k}_{\perp})=A \exp \left[-\frac{1}{8 \beta^{2}} \frac{\mathbf{k}_{\perp}^{2}+m^{2}}{x(1-x)}\right].
$$
As there are ambiguities in extending the non-relativistic form wave function
into a relativistic one, there are other possible prescriptions for
the transformed light-cone momentum space wave function~\cite{Terentev:1976jk}
$$
\psi\left(x, \mathbf{k}_{\perp}\right)=A \sqrt{\frac{1}{2 x(1-x)}} \exp \left[-\frac{m^{2}+\mathbf{k}_{\perp}^{2}}{8 \beta^{2} x(1-x)}\right],
$$
where the factor $\sqrt{1 / 2 x(1-x)}$ arises from the Jacobian factor.

\subsection{The holography prescription}
The light-cone holographic QCD is based on the correspondence~\cite{Maldacena:1997re,Gubser:1998bc,Witten:1998qj} between string states defined on the five-dimensional anti-de Sitter (AdS) space-time and conformal field theories (CFT) in physical spacetime. Although the conformal symmetry of the classical QCD Lagrangian with massless quarks is broken by quantum effects, the AdS/CFT correspondence can be used to obtain a first approximation to QCD at high energy or short distance because of its asymptotic freedom~\cite{Gross:1973ju,deAlfaro:1976vlx}.

We hope to find a semiclassical approximation of the relativistic wave function for strongly coupled QCD to calculate the meson bound state. To find such a differential equation, we need to simplify the light-cone Hamiltonian eigenvalues matrix equation to a valid light-cone Schrödinger equation. Because the kinematical and dynamical terms in QCD light-cone Hamiltonian is separated, we can express the mass square of a hadron in terms of light-cone wave functions as
$$
\begin{aligned}
M_{H}^{2}=& \int[d x]\left[d^{2} \mathbf{k}_{\perp}\right] \sum_{i=1}^{n} \frac{\mathbf{k}_{\perp i}^{2}+m_{i}^{2}}{x_{i}}\left|\psi\left(x_{i}, \mathbf{k}_{ \perp i}\right)\right|^{2} \\
&+\int[d x]\left[d^{2} \mathbf{k}_{\perp}\right] \psi^{*}\left(x_{i}, \mathbf{k}_{\perp i}\right) U \psi\left(x_{i}, \mathbf{k}_{ \perp i}\right),
\end{aligned}
$$
where $U$ is an effective potential. Using the Fourier transformation, one may obtain the expression in the coordinate space. Transforming into cylindrical coordinate system and separate variables~\cite{Brodsky:2007hb}, one obtains 
$$
\psi\left(x, \mathbf{b}_{\perp}\right)\longrightarrow e^{i L \varphi} X(x) \frac{\phi(\zeta)}{\sqrt{2 \pi \zeta}}.
$$
For a two-body system, the eigenequation with massless constituents in the cylindrical coordinate system is expressed as~\cite{deTeramond:2008ht}
$$
\left(-\frac{d^{2}}{d \zeta^{2}}-\frac{1-4 L^{2}}{4 \zeta^{2}}+\tilde{U}\right) \phi(\zeta)=M_{H}^{2} \phi(\zeta).
$$
$\mathbf{b}_{\perp}$ is the Fourier conjugate of $\mathbf{k}_{\perp}$. $\zeta=\sqrt{x(1-x)}\left|\mathbf{b}_{\perp}\right|$ is Lorentz invariant, and it measures the separation between the two quarks. It corresponds to the holographic variable $z$ in AdS space.

One can get the effective potential from the deformation of the AdS space~\cite{Ballon-Bayona:2021ibm}. The propagation of massive scalar modes in AdS space is described by the normalized solutions to the wave equation~\cite{Brodsky:2007hb}
$$
\left(-z^{2} \partial_{z}^{2}+\left(k-z \varphi^{\prime}(z)\right) z \partial_{z}+(m R)^{2}-z^{2} M^{2}\right) \Phi(z)=0,
$$
where $\varphi(z)$ is a function of the holographic coordinate $z$ called dilaton
which vanishes in the ultraviolet limit $z\rightarrow0$. The Regge behavior for bayrons requires $\varphi$ to have the form $\kappa^{2} \zeta^{2}$~\cite{Brodsky:2008pg,Abidin:2009hr}.

One can get the precise relation between $z$ and $\zeta$ by comparing the expression of the electromagnetic form factor in the AdS space and the physical space. For a two-parton bound state, in soft-wall model, the precise relation is given by~\cite{Brodsky:2007hb,Brodsky:2014yha}
$$
\begin{aligned}
\Phi_{L}(z) & = \left(\frac{R}{z}\right)^{L-\frac{d-1}{2}} e^{\varphi(z) / 2} \phi_{L}(z) ,\\
X(x) & = \sqrt{x(1-x)}.
\end{aligned}
$$
By this relation and the equation of $z$ in AdS space one can finally get
$$
\left[-\frac{d^{2}}{d \zeta^{2}}+V(\zeta)\right] \phi(\zeta)=M^{2} \phi(\zeta),
$$
and 
$$
V(\zeta)=-\frac{1-4 L^{2}}{4 \zeta^{2}}+\kappa^{4} \zeta^{2}+2 \kappa^{2}(L-1),
$$
for the valence state.
We can get effective light-cone wave function for a two-parton $L=0$ ground state in impact space as~\cite{Brodsky:2007hb}
$$
\psi_{q \bar{q} / \pi}\left(x, \mathbf{b}_{\perp}\right)\propto \kappa x(1-x) e^{-\frac{1}{2} \kappa^{2} x(1-x) \mathbf{b}_{\perp}^{2}}.
$$
The Fourier transformation of wave founction in momentum space is
$$
\psi_{\bar{q} q / \pi}\left(x, \mathbf{k}_{\perp}\right) \propto\frac{1}{\sqrt{x(1-x)}} e^{-\frac{\mathbf{k}_{\perp}^{2}}{2 \kappa^{2} x(1-x)}}.
$$
The mass of light quarks can be introduced by this substitution~\cite{Brodsky:2014yha,Liu:2015jna}
$$
\mathbf{k}_{\perp}^{2}\rightarrow \mathbf{k}_{\perp}^{2}+m^{2},
$$
then 
$$
\psi_{\bar{q} q / \pi}\left(x, \mathbf{k}_{\perp}\right) =\frac{A}{\kappa\sqrt{x(1-x)}} e^{-\frac{\mathbf{k}_{\perp}^{2}+m^{2}}{2 \kappa^{2} x(1-x)}}.
$$
If we define $\frac{A}{\kappa}=A^{\prime}$, then
$$
\psi\left(x, \mathbf{k}_{\perp}\right)=\frac{A^{\prime}}{ \sqrt{x(1-x)}} e^{-\frac{\mathbf{k}_{\perp}^{2}+m^{2}}{2 \kappa^{2} x(1-x)}}.
$$
We can find that it has the same form as the BHL prescription with a Jacobian factor, and there are relationships among the parameters
$$
\begin{aligned}
A & = A^{\prime }, \\
2\kappa^{2} =8\beta ^{2} &\to  \kappa = 2\beta .
\end{aligned}
$$

This may imply that using the soft-wall model in AdS space and choosing a dilaton of $\kappa^{2} z^{2}$ to break the symmetry of the AdS space to introduce a kinetic effect are equivalent to using a harmonic oscillator potential in the light-cone form to approximate the kinetic effect between two quarks.
\section{Numerical calculations}
The specific expression of $F_{\gamma \gamma^{*} \rightarrow \pi}\left(Q^{2}\right)$ as giving in Sec.~II is
$$
F_{\gamma \gamma^{*} \rightarrow \pi}\left(Q^{2}\right)=4 \sqrt{3}\left(e_{u}^{2}-e_{d}^{2}\right) \int_{0}^{1} \mathrm{~d} x \int \frac{\mathrm{d}^{2} \mathbf{k}_{\perp}}{16 \pi^{3}} \psi_{\pi}\left(x, \mathbf{k}_{\perp}^{\prime}\right) \frac{m}{x \sqrt{m^{2}+\mathbf{k}_{\perp}^{\prime 2}}} \frac{x(1-x)}{m^{2}+\mathbf{k}_{\perp}^{2}},
$$
in which $\mathbf{k}_{\perp}^{\prime}=\mathbf{k}_{\perp}+(1-x) \mathbf{q}_{\perp}$, $\psi_{\pi}\left(x, \mathbf{k}_{\perp}^{\prime}\right)$ is the momentum space wave function. It is more convenient to switch to the polar coordinate system for numerical calculations

$$
\begin{aligned}
F_{\gamma \gamma^{*} \rightarrow \pi}\left(Q^{2}\right)=&4 \sqrt{3}\left(e_{u}^{2}-e_{d}^{2}\right) \int_{0}^{1} \mathrm{~d} x \int \frac{\mathbf{k}_{\perp}\mathrm{d} \mathbf{k}_{\perp}\mathrm{d}\theta }{16 \pi^{3}} \psi_{\pi}\left(x, \mathbf{k}_{\perp}\right)
\frac{m}{x \sqrt{m^{2}+\mathbf{k}_{\perp}^{ 2}}}\\ &\frac{x(1-x)}{m^{2}+\mathbf{k}_{\perp}^{2}+Q^{2}(1-x)^{2}-2Q\left | \mathbf{k}_{\perp} \right | (1-x)\mathrm{cos}\theta }.
\end{aligned}
$$

Now we need to determine the form of the momentum wave function. According to the previous discussion, we have three forms.

The BHL prescription
$$
\psi(x, \mathbf{k}_{\perp})=A \exp \left[-\frac{1}{8 \beta^{2}} \frac{\mathbf{k}_{\perp}^{2}+m^{2}}{x(1-x)}\right].
$$
The BHL prescription with Jacobian factor
$$
\psi(x, \mathbf{k}_{\perp})=\frac{{A}}{\sqrt[]{x(1-x)} } \exp \left[-\frac{1}{8 \beta^{2}} \frac{\mathbf{k}_{\perp}^{2}+m^{2}}{x(1-x)}\right].
$$
The holographic QCD approach
$$
\psi\left(x, \mathbf{k}_{\perp}\right)=\frac{A}{\kappa \sqrt{x(1-x)}} e^{-\frac{\mathbf{k}_{\perp}^{2}+m^{2}}{2 \kappa^{2} x(1-x)}}.
$$
This is consistent with the BHL prescription considering the Jacobian factor, so we only use the first two methods in our calculation.

These wave functions all contain some independent parameters. Common constraints are

1. The first and most natural limitation is the normalization of the wave function. Since we only consider the valence state, the normalization of the wave function should be less than or equal to one
$$
\int_{0}^{1} d x \int \frac{d^{2} \mathbf{k}_{\perp}}{16 \pi^{3}}\psi^{*}_{\pi}\left(x, \mathbf{k}_{\perp}\right) \psi_{\pi}\left(x, \mathbf{k}_{\perp}\right)\le 1.
$$
2.The $\pi^{+}$ weak decay constant$f_{\pi}=92.4~\mathrm{MeV}$ ~\cite{ParticleDataGroup:2020ssz} is defined by
$$
\left\langle0\right| \bar{u} \gamma^{+}\left(1-\gamma_{5}\right) d\left | \pi \right \rangle =-\sqrt{2} f_{\pi} p^{+}.
$$
Paper~\cite{Xiao:2003wf} gives that
$$\frac{f_{\pi}}{2 \sqrt{3}}=\int_{0}^{1} d x \int \frac{d^{2} \mathbf{k}_{\perp}}{16 \pi^{3}} \frac{\left(k_{1}^{+}+m\right)\left(k_{2}^{+}+m\right)-\mathbf{k}_{\perp}{ }^{2}}{\left[\left(k_{1}^{+}+m\right)^{2}+\mathbf{k}_{\perp}{ }^{2}\right]^{1 / 2}\left[\left(k_{2}^{+}+m\right)^{2}+\mathbf{k}_{\perp}{ }^{2}\right]^{1 / 2}} \psi_{\pi}\left(x, \mathbf{k}_{\perp}\right),$$
and one can simplify the above formula with $q_{i}^{+}=x_{i} M$ and $M=\left(m^{2}+\mathbf{k}_{\perp}^{2}\right) / x(1-x)$ 
$$
\frac{f_{\pi}}{2 \sqrt{3}}=\int_{0}^{1} d x \int \frac{d^{2} \mathbf{k}_{\perp}}{16 \pi^{3}} \frac{m}{\sqrt{m^{2}+\mathbf{k}_{\perp}^{2}}} \psi_{\pi}\left(x, \mathbf{k}_{\perp}\right).
$$
3.The charged mean square radius of $\pi^{+}$ is related to electromagnetic form factor with~\cite{Dally:1982zk}
$$
\left\langle r_{\pi^{+}}^{2}\right\rangle=-\left.6 \frac{\partial F_{\pi^{+}}\left(Q^{2}\right)}{\partial Q^{2}}\right|_{Q^{2}=0}.
$$
4. Brodsky and Lepage give an asymptotic behavior of $Q^{2} \longrightarrow \infty $ by perturbative QCD~\cite{Brodsky:1981rp}
$$\lim _{Q^{2} \rightarrow \infty} F_{\pi  \rightarrow \gamma \gamma^{*}}\left(Q^{2}\right)=\frac{2  f_{\pi}}{Q^{2}}.$$
5. The decay width of $\pi^{0}$~\cite{CLEO:1997fho} has the following relationship with TFF
$$
\left|F_{\gamma \gamma^{*} \rightarrow \pi}(0)\right|^{2}=\left|F_{\pi \rightarrow \gamma \gamma}(0)\right|^{2}=\frac{64 \pi \Gamma\left(\pi^{0} \rightarrow \gamma \gamma\right)}{(4 \pi \alpha)^{2} M_{\pi}^{3}}.
$$

With the BHL wave function, paper~\cite{Xiao:2003wf} calculates the $\pi$ TFF through the constraints 2,3 and 5; paper~\cite{Xiao:2005af,Qian:2008px} calculates the $\eta$ and $\eta^{\prime}$ TFF through the constraints 3,4 and 5. Although these works agree well with the experiment, the normalization of these wave functions is greater than one. We should limit these parameters with processes that are only related to valence states~\cite{Huang:1994dy}. The electromagnetic radius should be related to sea quarks, so it is not a good condition. Our model may not be suitable for high energy scales, and according to the experimental data, the behavior of the TFF at high energy scales is also inconsistent with this asymptotic behavior predicted by perturbative QCD.

We choose these conditions
$$
\begin{aligned}
f_{\pi}  &= 92.4 ~\mathrm{MeV}, \\
F_{\pi \rightarrow \gamma \gamma}(&0)  = 0.271~ \mathrm{GeV}^{-1}, \\
\int_{0}^{1} d x \int \frac{d^{2} \mathbf{k}_{\perp}}{16 \pi^{3}}&\psi^{*}_{\pi}\left(x, \mathbf{k}_{\perp}\right) \psi_{\pi}\left(x, \mathbf{k}_{\perp}\right)  = 1,
\end{aligned}
$$
to constrain these parameters.

The BHL prescription, $m=0.196 ~\mathrm{GeV}$, $\beta=0.428 ~\mathrm{GeV}$, $ A=41.53 ~\mathrm{GeV}^{-1}$. We can calculate above constraints with these parameters
$$
\begin{aligned}
f_{\pi} &=86.35~ \mathrm{MeV}, \\
 F_{\pi \rightarrow \gamma \gamma}(0) &=0.252 ~\mathrm{GeV}^{-1}.\\
\end{aligned}
$$

The holography prescription and the BHL prescription with Jacobian factor are formally identical, and we choose the latter in our calculation. It gives $m=0.197 ~\mathrm{GeV}$, $\beta=0.447 ~\mathrm{GeV}$, $ A=17.048 ~\mathrm{GeV}^{-1}$, with these parameters,  
$$
\begin{aligned}
f_{\pi} &=89.82~ \mathrm{MeV}, \\
 F_{\pi \rightarrow \gamma \gamma}(0) &=0.262 ~\mathrm{GeV}^{-1}.\\
\end{aligned}
$$
The two wave functions are all normalized to 1.

It can be seen that the $\sqrt{x(1-x)}$ factor can improve the fitting of the constraints under the normalization conditions. We find that the calculated results using the normalized wave function are slightly better than the unnormalized results below 1~GeV to compare with the experimental results.

With these parameters to calculate the transition form factor, it can be seen from Figure 3 that the calculated results using the parameters obtained by considering the normalized fitting are in better agreement with the experimental results below $1~\mathrm{GeV}$.
\begin{figure}[H]
    \centering
    \includegraphics[width=0.8\textwidth]{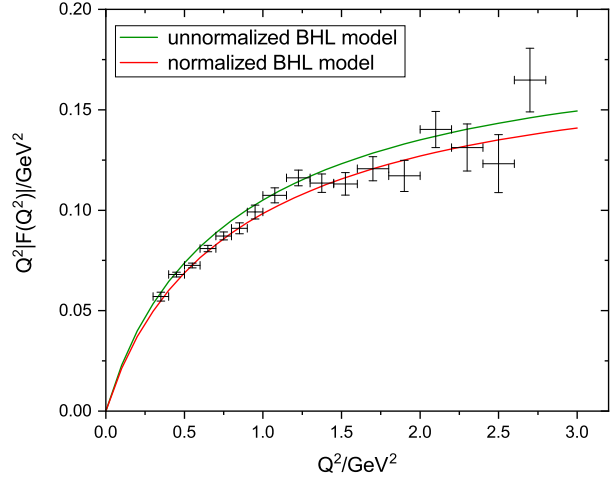}
    \caption{Comparison of the calculated results of the parameters fitted with the normalized condition and the calculated results of the unnormalized wave function~\cite{Xiao:2003wf} with the preliminary experimental data at BESIII~\cite{Redmer:2019zzr}.}
    \label{nom}
\end{figure}

The calculation results shown in Fig.4 show that both two schemes are in good agreement with the experimental results, especially in the energy scale below $1~\mathrm{GeV}$.
\begin{figure}[H]
    \centering
    \includegraphics[width=0.8\textwidth]{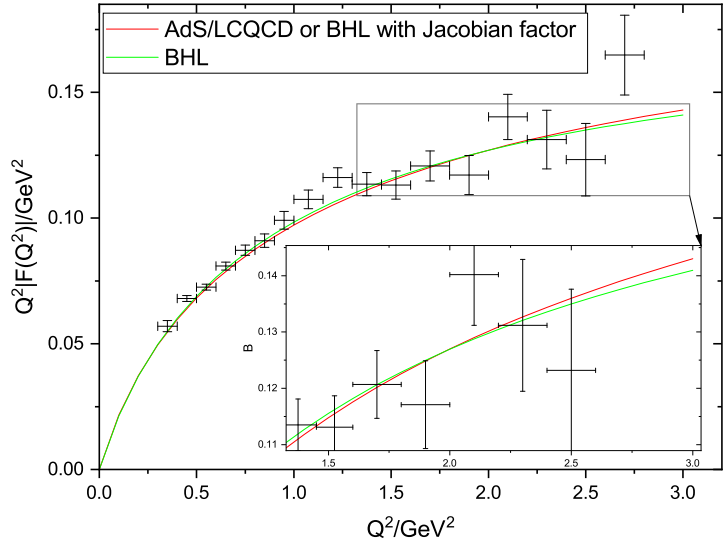}
    \caption{Comparison of the $\pi^{0}$ TFF calculated by the two models with the preliminary results at BESIII~\cite{Redmer:2019zzr}.}
    \label{JS}
\end{figure}

\section{Comparison with other theoretical results}
There have been some previous works on hadron properties using light-cone quantum field theory. 
Xiao and Ma calculated $\pi^{0}$ TFF~\cite{Xiao:2003wf}, and Qian and Ma calculated $\eta,\eta^{\prime}$ TFF~\cite{Qian:2008px}. 
There are also some works on properties of other mesons and baryons~\cite{Ma:2002ir,Xiao:2002iv,Liu:2014npa}. The main difference between these theoretical works is that the conditions that limit the parameters in the model and the description of interactions between quarks are different. Some works give those parameters by directly fitting the TFF experimental data. We have pointed out in Sec.~5 that the constraints we chose are reasonable in the energy scales we study.

There are many other theories on photon-pion TFF besides light-cone quantum field theory, and we make a briefly review as follows. The first is the phenomenological model. Many phenomenological theories have been proposed to describe the photon-pion TFF. The simplest is the vector meson dominance~(VMD) model~\cite{Meissner:1987ge}, for the double-virtual case
$$
F_{\pi^{0} \gamma^{*} \gamma^{*}}^{V M D}\left(Q_{1}^{2}, Q_{2}^{2}\right)=\frac{\alpha M_{V}^{4}}{\left(M_{V}^{2}-Q_{1}^{2}\right)\left(M_{V}^{2}-Q_{2}^{2}\right)},
$$
in which $\alpha =1 /\left(4 \pi^{2} f_{\pi}\right)$ to satisfy the chiral anomaly, and $M_{V}$ is the mass of vector meson $\rho$. In addition, there are lowest meson dominance~(LMD) models~\cite{Peris:1998nj}, or models that combine VMD with LMD~\cite{Knecht:2001xc}. These models meet more experimental or theoretical constraints by introducing more parameters. Although phenomenological models are convenient to describe TFF, they do not reflect the full behaviors of form factors and may be unreliable in some energy regions.

Obviously phenomenological models are not enough, and we need to understand TFF at a more fundamental level. The basic theory of hadron is QCD. Due to the non-perturbative characteristics of QCD in the energy scale we study, most of the works that tried to study TFF in perturbative theory~\cite{delAguila:1981nk,Braaten:1982yp} are not very satisfactory until two loops~\cite{Gao:2021iqq,Gao:2021beo}.

Integral equations and lattice QCD are also methods for non-perturbative calculations of hadron properties besides light-cone quantum field theory. The Dyson-Schwinger equation~(DSE) based on QCD can realize various non-perturbative properties of QCD~\cite{Roberts:1994dr}. Further combined with the Bethe-Salpeter equation, the properties of mesons can be studied~\cite{Eichmann:2017wil}. The integral equation methods are limited by the rainbow truncation. Lattice QCD is a method of computing hadrons from first principles without perturbations, though it requires a lot of computing resources. 
Gérardin, Meyer and Nyffeler calculated the photon-pion TFF using the lattice QCD~\cite{Gerardin:2016cqj,Gerardin:2019vio}. 
They used the LMD model when extracting TFF from lattice calculations, although they claimed that their results are model-independent. Some people are trying to do a lattice calculation that does not introduce any models at all. Limited by computational resources, the err of this kind of works is large in the energy range that can be compared with the experiment at present. The comparison of the lattice QCD calculation~\cite{Gerardin:2019vio} and DSE approach~\cite{Eichmann:2017wil} with our work is shown in Fig.~5.

\begin{figure}[H]
    \centering
    \includegraphics[width=1.0\textwidth]{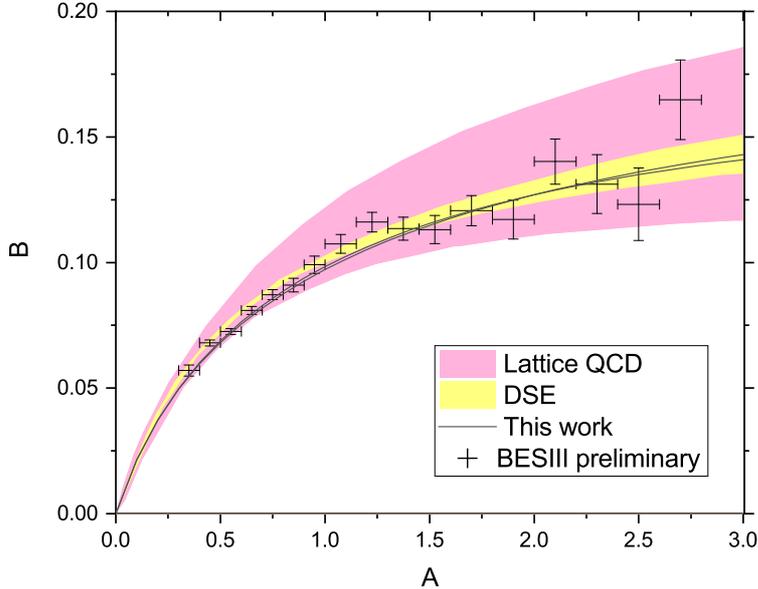}
    \caption{Comparison of BSEIII preliminary data~\cite{Redmer:2019zzr} with calculated results from Lattice QCD~\cite{Gerardin:2019vio}, DSE approach~\cite{Eichmann:2017wil} and this work.}
    \label{db}
\end{figure}

\section{Conclusion}
The light-cone formalism provides a convenient framework for the relativistic description
of hadrons in terms of quark and gluon degrees of freedom, and for the application of
perturbative QCD to exclusive processes. With light-cone quark model, we find that our numerical prediction for the $\gamma \gamma^{*}\longrightarrow\pi$
transition form factor agrees well with the latest BESIII preliminary experimental data at the energy scale below 3~GeV. We derive the momentum space wave function in light-cone formalism by AdS/LCQCD mapping and consider light quark mass by a substitution. We found that the wave function obtained in this way has the same form as the BHL wave function with the Jacobian factor. This implies that in the soft-wall model, the dynamic effect introduced by breaking the symmetry of AdS space through the dilatons of the form $\kappa^{2}z^{2}$ is equivalent to taking the harmonic oscillator potential in the physical space. Both the two models agree well with the BESIII preliminary experimental data. This shows that the assumption of harmonic oscillator potential is reasonable and the breaking of the conformal symmetry by quark mass is relatively little at this scale. 

\section*{Declaration of Competing Interest}
The authors declare that they have no conflicts of interest in this
work.

\section*{Acknowledgements}

This work is supported by National Natural Science Foundation of China (Grant No.~12075003).

\bibliographystyle{elsarticle-num}

\bibliography{sample}
\end{document}